\documentclass[aps,twocolumn,a4paper,showpacs,showkeys]{revtex4}
\usepackage{graphicx}
\usepackage{amsmath}
\usepackage{amssymb}

\begin{document}

\title{Numbers of $n$-th neighbors and node-to-node distances in growing networks\footnote{Dedicated to 
Professor Andrzej Z. Maksymowicz on the occasion of his 65th birthday}}

\author{Krzysztof Malarz}
\homepage{http://home.agh.edu.pl/malarz/}
\affiliation{
Faculty of Physics and Applied Computer Science,
AGH University of Science and Technology,\\
al. Mickiewicza 30, PL-30059 Krak\'ow, Poland
}

\date{\today}

\begin{abstract}
Topology of exponential and scale-free trees and simple graphs is investigated numerically. 
The numbers of the nearest neighbors, the next-nearest neighbors, the next-next-nearest neighbors, the 4-th and the 5-th neighbors are calculated.
The functional dependence of the node-to-node distance $d_{ij}$ on the product of connectivities $k_ik_j$ has been also checked.
The results of simulations for exponential networks agree with the existing analytical predictions. 
\end{abstract}



\maketitle

%% ######################################################
\section{Introduction}
%% ######################################################
Complex networks have been attracting great attention for decades.
They may describe many real-world systems in 
the social sciences, biology, computer science, telecommunication and others \cite{newman-rev,d-m-rev}.
Mathematical description of networks is provided by the graph theory \cite{graph}. 
Graph is a set of vertexes (nodes) connected by edges (links).
The main local characteristic of a graph is the node degree, i.e. the number of links incoming to or outgoing from a node.
For almost fifty years, the paradigm of ``typical node'' has been present in the science of networks.
Networks of typical nodes were described by Erd\H os and R\'enyi \cite{crg} ({\em classical random graphs} --- CRG).
In their model, $N$ nodes are connected randomly with $L$ links: each inter-node link is realized with the probability $p=2L/[N(N-1)]$.
In this model, the node degree distribution is given by a Poisson law, 
i.e. $P_k(k)=\exp(-\{ k \})\{ k \}^k/k!$, where $\{\cdots\}$ denotes the mean over all $N$ nodes, and the node degrees observed on a CRG fluctuate around $\{ k \}$.

As pointed out by Albert and Barab\'asi in their seminal paper \cite{ab-org}, networks in real world more often exhibit a power-like degree distribution, 
i.e. $P_k(k)\propto k^{-\gamma}$.
In the Albert--Barab\'asi (A-B) model, the node degrees assume all integer values in the thermodynamic limit and there is no characteristic value of the degree.
Thus, with this observation the Hungarian mathematicians' world of networks with typical nodes became a world of scale-free networks.

CRG and A-B networks are two examples belonging to two different families of networks \cite{waclaw}.
The first one belongs to the so-called {\em homogeneous} networks, which may be described via a statistical ensemble. 
The A-B networks have temporal structure, as they come into being via growth process.
The A-B network is an example of {\em causal} network.

For networks, the act of growing means subsequent attachments of new nodes, each with $M$ links, to previously existing nodes.
The procedure of selection of those ``old'' nodes influences the network topology and the degree distribution.
When old nodes are selected randomly --- i.e. the probability of attachment is the same for all nodes ---  
{\em exponential} networks are created and the nodes degree distribution is an exponential one \cite{d-m-rev}.
On the other hand, when the attachment is preferential --- i.e. the probability of choosing a node is proportional 
to its degree --- the degree distribution is power-like and network can be termed as {\em scale-free} \cite{newman-rev}.

The number of edges $M$ also influences the network topology:
\begin{itemize}
\item when $M=1$, the path between any pair of nodes is unique; the growing structure is called {\em a tree},
\item when $M>1$, cyclic paths are possible and {\em graph} looses its tree-like properties,
\item when $M>1$ and chosen old nodes all are different, multiple edges in the network are absent and the structure is {\em a simple graph}.
\end{itemize}
Such attaching procedure prevents possibility of {\em loops}, i.e. self-links.

Several characteristic of real or simulated networks may be practically useful.
For example many papers discuss the networks resistance to possible damage \cite{damage}, 
their tolerance on random and/or intentional attack \cite{attack} or transport properties in terms either of
the percolation theory \cite{percol} or of the shortest path finding \cite{epjb,short}.
Newman {\em et al.} applied the generating function formalism \cite{genfun} to evaluate the number of nodes 
%% ------------------------------------------------------
\begin{equation}
\label{eq-zm}
z_m=  {z_1}^{2-m} {z_2}^{m-1}
\end{equation}
%% ------------------------------------------------------
in subsequent ($m$-th) layer from a randomly chosen origin \cite{newman}.
In Eq. \eqref{eq-zm} $z_1$ and $z_2$ are typical values of the number of nodes nearest neighbors and the number of nodes next-nearest neighbors, respectively.
The first one ($z_1$) is obviously equal to average node degree $z_1=\{ k_1\}$.
The latter ($z_2$) was evaluated lately by Shargel {\em et al.} \cite{shargel} as 
%% ------------------------------------------------------
\begin{equation}
\label{eq-z2}
z_2=\{k^2\}-\{k\}.
\end{equation}
%% ------------------------------------------------------

Basing on the same technique, Motter {\em et al.} \cite{motter} derived the average node-to-node distance $d_{ij}$ 
dependence on the product $k_ik_j$ of the node degrees for random networks:
%% ------------------------------------------------------
\begin{equation}
\label{eq-holyst}
\langle d_{ij} \rangle = A-B\ln(k_ik_j),
\end{equation}
%% ------------------------------------------------------
where $\langle\cdots\rangle$ denotes the average over all node pairs, the product of the pair degrees being equal to $k_ik_j$.
Recently, Ho{\l}yst {\em et al.} \cite{holyst} have confirmed this dependence numerically and presented some examples of real-world networks which obey Motter {\em et al.} theoretical predictions.

In this paper we check if Motter {\em et al.}, Ho{\l}yst {\em et al.} and Shargel {\em et al.} predictions apply to the growing exponential networks.
Namely, we evaluate number of neighbors in subsequent layers.
The node-to-node distance vs. product of their degrees is also simulated.
For completeness, the calculations and discussion include the growing scale-free A-B networks.

In the next section we explain our numerical approach.
In section \ref{sec-res} we present results of Monte Carlo simulations of the average number of subsequent neighbors (\ref{sec-zm}) and the inter-nodes distance dependence on the product of their degrees (\ref{sec-dis}).
The last section summarizes the results.

%% ######################################################
\section{Numerical approach}
%% ######################################################
Numerical approach is based on an ``on-line'' construction of the distance matrix $\mathbf{D}$ during the network growth \cite{ijmpc,task,physicaa,app}.
An element $d_{ij}$ of the distance matrix  gives the length of the shortest path between nodes $i$ and $j$, i.e.  the minimal number of edges which connect these vertexes.
The numbers $d_{ij}$ in $i$-th row/column inform how far is the node $i$ from another node $j$.
Then, the number $z_m(i)$ representing the number of occurrences of the $m$ value in the $i$-th matrix row/column gives the information how many neighbors of the node $i$ are at the distance $m$ \cite{app}.
The average number of the matrix elements of a given value in all rows/columns --- i.e. in the whole matrix --- gives a typical number $z_m$ of subsequent neighbors, for example, the nearest neighbors for $m=1$, the next-nearest neighbors for $m=2$, the next-next-nearest neighbors for $m=3$, etc. 
Additionally, the number of unities (``1'') in the $i$-th row/column gives degree of the $i$-th node: $k(i)=z_1(i)$.

%% ######################################################
\section{Results of simulations}
\label{sec-res}
%% ######################################################

We construct the distance matrix $\mathbf{D}$ for $N=10^3$ nodes.
The results are averaged over $N_{\text{run}}=10^4$ independent simulations.

%% ======================================================
\subsection{Number of nodes in subsequent layers}
\label{sec-zm}
\begin{figure*}
\begin{center}
\includegraphics[scale=.6]{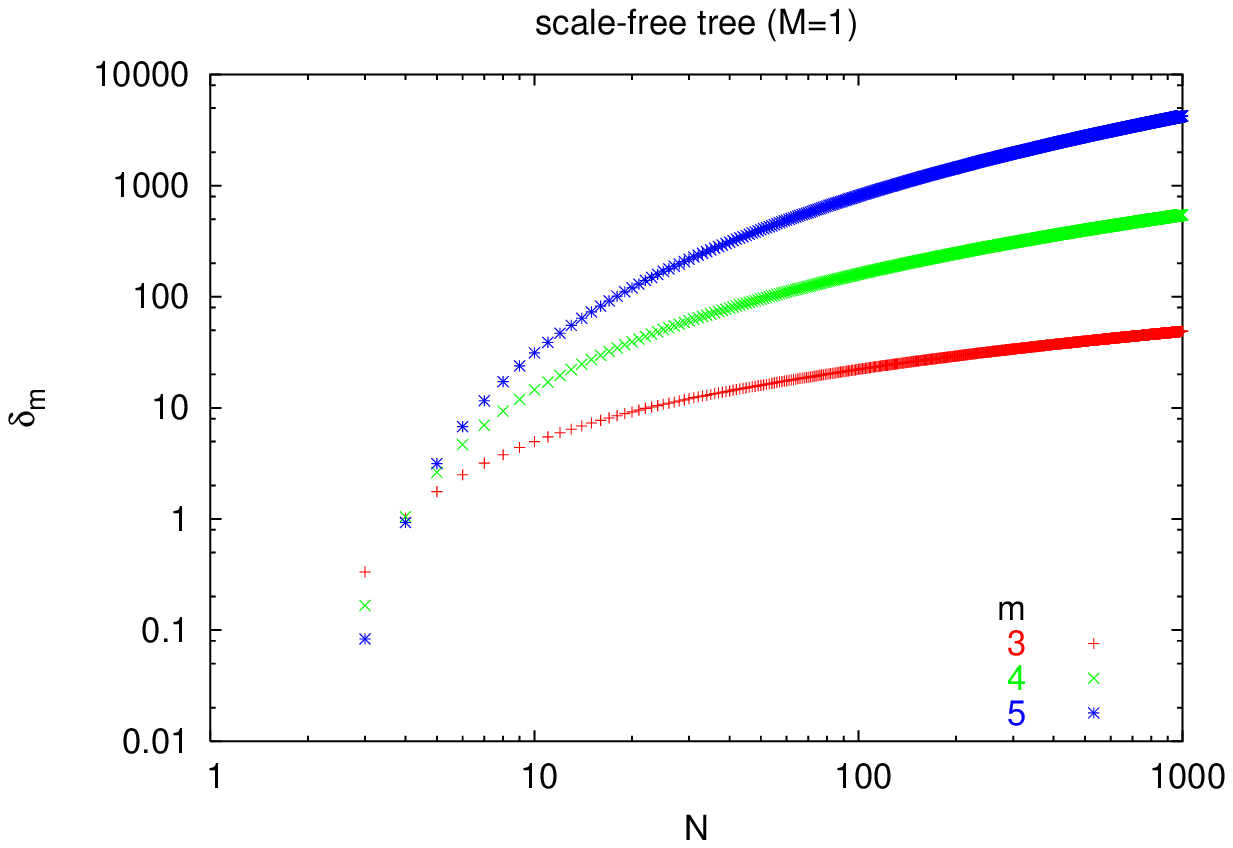}
\includegraphics[scale=.6]{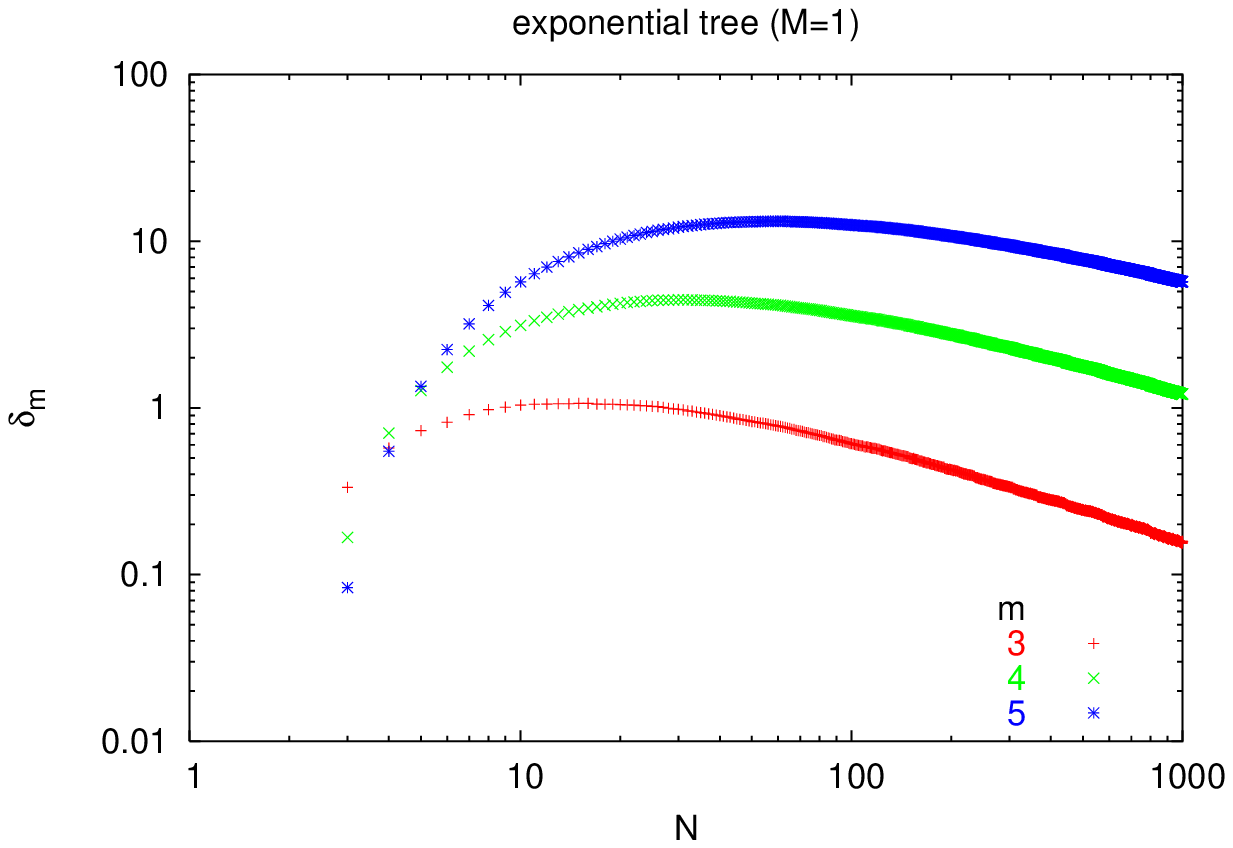}\\
\includegraphics[scale=.6]{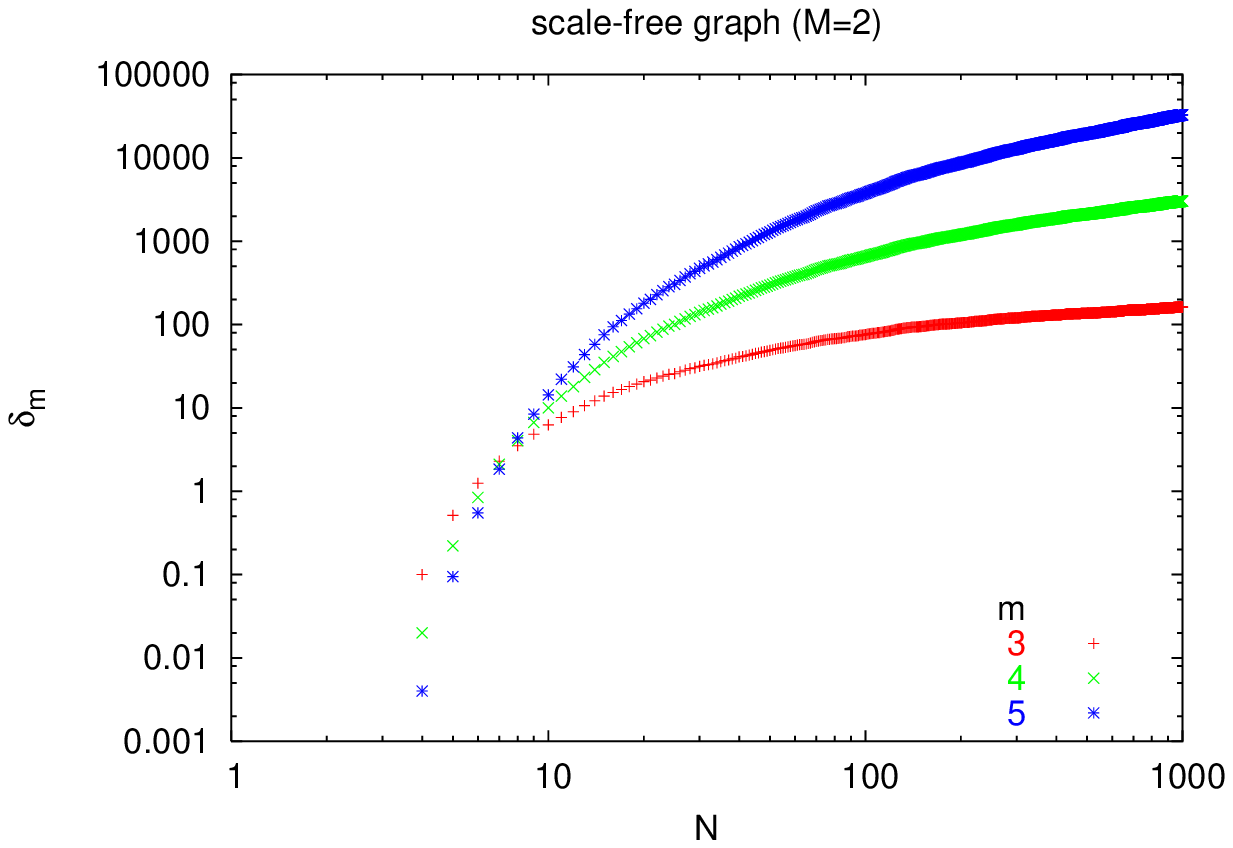}
\includegraphics[scale=.6]{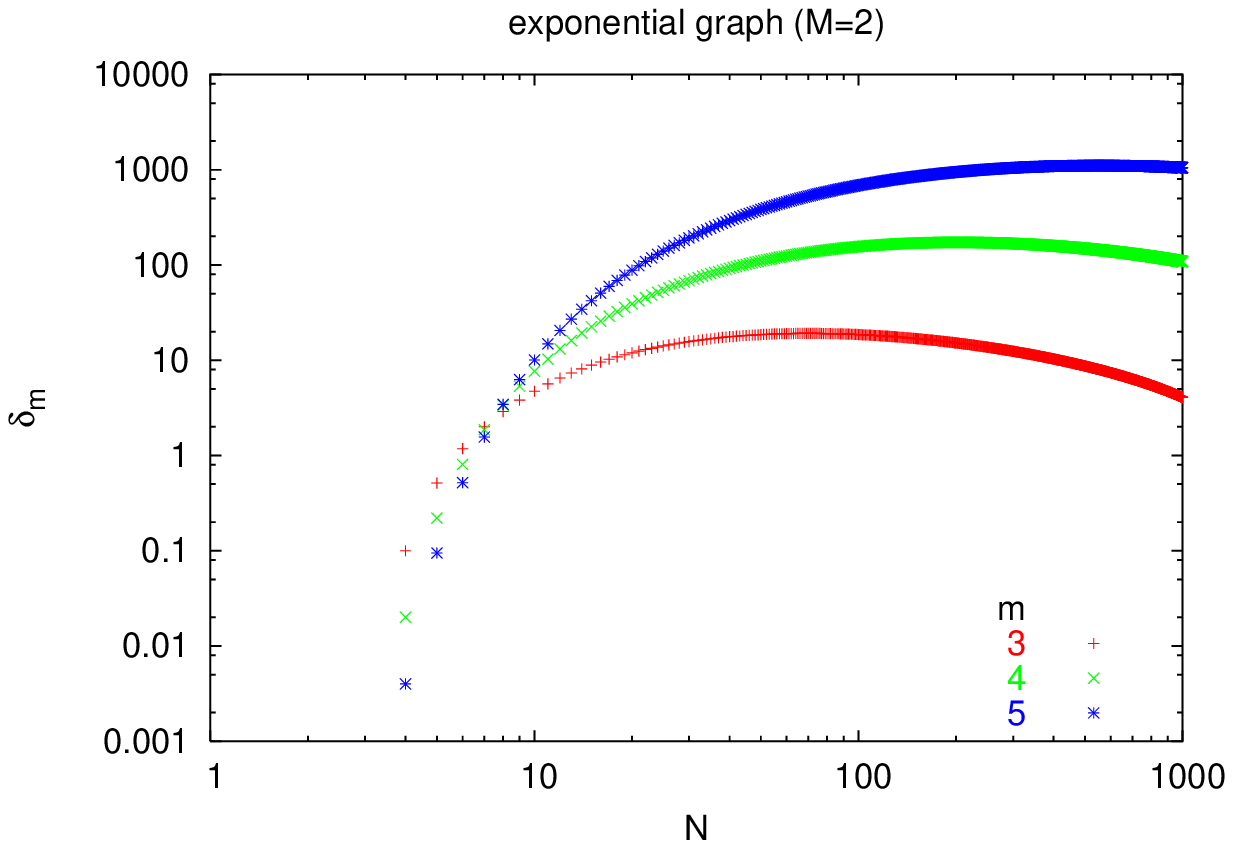}
\end{center}
\caption{\label{fig-z} Dependence $\delta_m = {z_1}^{2-m} {z_2}^{m-1} - z_m$ ($m=3,4,5$) on network size $N$ for growing exponential and scale-free trees ($M=1$) and simple graphs ($M=2$).}
\end{figure*}

%% ======================================================
Fig. \ref{fig-z} shows how the deviation $\delta_m\equiv z_1^{2-m}z_2^{m-1}-z_m$ between $z_m$ calculated from Eq. \eqref{eq-zm} and from the direct simulation behaves as the function of the system size $N$ for $m=3,4,5$.
%% Additionally, theoretically predicted \cite{newman} values $z_m$ given by Eq. \eqref{eq-zm} are included.
As one can see, starting with $N\approx 100$ this difference decreases with $N$ for the exponential networks.
For the scale-free networks either the number of nodes $(N=10^3)$ is still too small to observe a good agreement between $z_m$ and $z_1^{2-m}z_2^{m-1}$ or Eq. \eqref{eq-zm} does not hold for the A-B graphs.

%% ------------------------------------------------------
%% ------------------------------------------------------

By construction, the average number of the nearest neighbors $z_1$ is 2 and 4 for trees and simple graphs, respectively.
The number of next-nearest neighbors $z_2$ depends on the applied rules of growth:
when the growth is governed by the preferential attachment rule, we have $z_2\approx 14$ and $z_2\approx 38$ 
for $M=1$ and $M=2$, respectively.
For the exponential networks, these numbers are $z_2\approx 4$ ($M=1$) and $z_2\approx 17$ ($M=2$).
As the average number of the nearest neighbors $z_1$ is exactly equal to the average nodes degree $\{k\}$, it may be evaluated from 
the degree distribution $P_k(k)$ as $\{k\}=\sum_{k=M}^\infty kP_k(k)$, as well.
For the exponential network this distribution \cite{d-m-rev,app} is given by 
%% ------------------------------------------------------
\begin{equation}
P_k(k\ge M)=
\begin{cases}
2^{-k}               & \text{for } M=1,\\
3/4 \cdot (3/2)^{-k} & \text{for } M=2,
\end{cases}
\end{equation}
while for the scale-free networks \cite{krapivsky00,pk-sfn} it is
\begin{equation}
P_k(k\ge M)=\dfrac{2M(M+1)}{(k+2)(k+1)k}.
\end{equation}
%% ------------------------------------------------------

\begin{figure}
\begin{center}
\includegraphics[scale=.6]{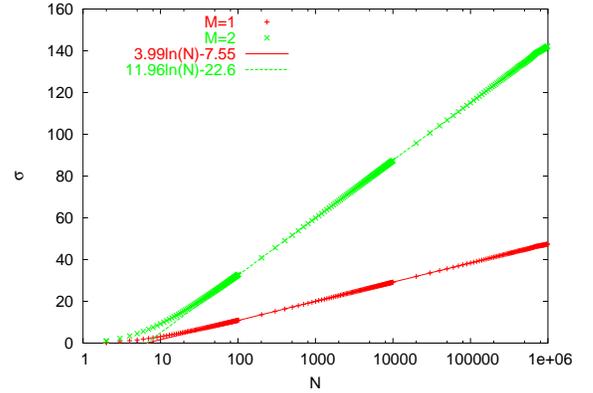}
\end{center}
\caption{\label{fig-k2-k} Dependence $\sigma(N)$ for growing scale-free networks.}
\end{figure}

The mean number of the next-nearest neighbors ($z_2$) may be evaluated basing on Eq. \eqref{eq-z2} which diverges for scale-free networks with $N\to\infty$.
For finite but large network this sum
%% ------------------------------------------------------
\begin{equation}
 \sigma\equiv \sum_{k=M}^{N-1} k(k-1)P_k(k) = 2M(M+1)\sum_{k=M}^{N-1} \dfrac{k-1}{(k+2)(k+1)}
\label{eq-sigma}
\end{equation}
%% ------------------------------------------------------
grows logarithmically with $N$, $\sigma=3.99\ln N-7.55$ ($M=1$) and $\sigma=11.96\ln N-22.6$ ($M=2$) as presented in Fig. \ref{fig-k2-k}. 

%% ------------------------------------------------------
%% ------------------------------------------------------

The results are collected in Tab. \ref{tab}.

%% ------------------------------------------------------
\begin{table}
\caption{\label{tab}
Average number of the nearest neighbors $z_1$ and the next-nearest neighbors $z_2$ for different evolving 
scale-free and exponential networks with $N=10^3$ nodes.
The results are averaged over $N_{\text{run}}=10^4$ samples.
Theoretical predictions of the average nodes degrees $\{k\} (=z_1)$ and $\{k^2\}-\{k\} (=z_2)$ are also included. 
Four last lines show the least-square fit coefficients $A$ and $B$ in the dependence 
$\langle d_{ij} \rangle = A-B\ln(k_ik_j)$ and their predictions $A^{\text{th}}$ and 
$B^{\text{th}}$ given by Eq. \eqref{eq-motter}.
}
\begin{ruledtabular}
\begin{tabular}{r cccc} 
      & \multicolumn{2}{c}{scale-free} & \multicolumn{2}{c}{exponential} \\
\hline     
$M$                           &  1       &   2      &   1     &   2     \\
$\sum_{k=M}^\infty kP_k$      &  2       &   4      &   2     &   4     \\
$z_1=\{k\}$                   &  1.998   &   3.994  &   1.998 &   3.994 \\
$\sum_{k=M}^\infty k(k-1)P_k$ & $\infty$ & $\infty$ &   4     &  18     \\
$\sum_{k=M}^{N-1} k(k-1)P_k$  & 19.95    &  59.86   &         &         \\ %% 999  19.9538937  59.8616791
$\{k^2\}-\{k\}$               & 13.68    &  39.66   &   3.966 &  17.81  \\
$z_2$                         & 13.68    &  38.11   &   3.966 &  17.37  \\
$z_2^2/z_1$                   & 93.6     & 363.6    &   7.88  &  75.6   \\
$z_3$                         & 44.8     & 201.5    &   7.72  &  71.5   \\
$A$                           &  7.68    &   5.09   &  12.8   &   6.77  \\
$B$                           &  0.783   &   0.438  &   1.73  &   0.746 \\
$A^{\text{th}}$               &  4.93    &   4.68   &  12.0   &   6.64  \\
$B^{\text{th}}$               &  0.519   &   0.443  &   1.46  &   0.679 \\
\end{tabular}
\end{ruledtabular}
\end{table}
%% ------------------------------------------------------

%% ======================================================
\subsection{Node-to-node distances and node degrees}
\label{sec-dis}
%% ======================================================
Using the generating function formalism \cite{genfun,newman} Motter {\em et al.} \cite{motter} derived an expression
for the length of the shortest path between the nodes for a given value of the product of connectivities $k_ik_j$:
%% ------------------------------------------------------
\begin{equation}
\label{eq-motter}
\begin{split}
\langle d_{ij} \rangle 
=\left[1+\dfrac{\ln(Nz_1)}{\ln(z_2/z_1)}\right]
-\left[\dfrac{1}{\ln(z_2/z_1)}\right]\ln(k_ik_j)\equiv\\
A^{\text{th}}-B^{\text{th}}\ln(k_ik_j).
\end{split}
\end{equation}
%% ------------------------------------------------------

Lately, such a kind of dependence $\langle d_{ij} \rangle$ vs $(k_ik_j)$ 
has been shown to be valid in few real-world networks, including biological and scientific papers citation networks,
public-transportation systems of several Polish towns, 
and simulated CRG and A-B networks \cite{holyst}.

%% ------------------------------------------------------
\begin{figure*}
\begin{center}
\includegraphics[scale=.60]{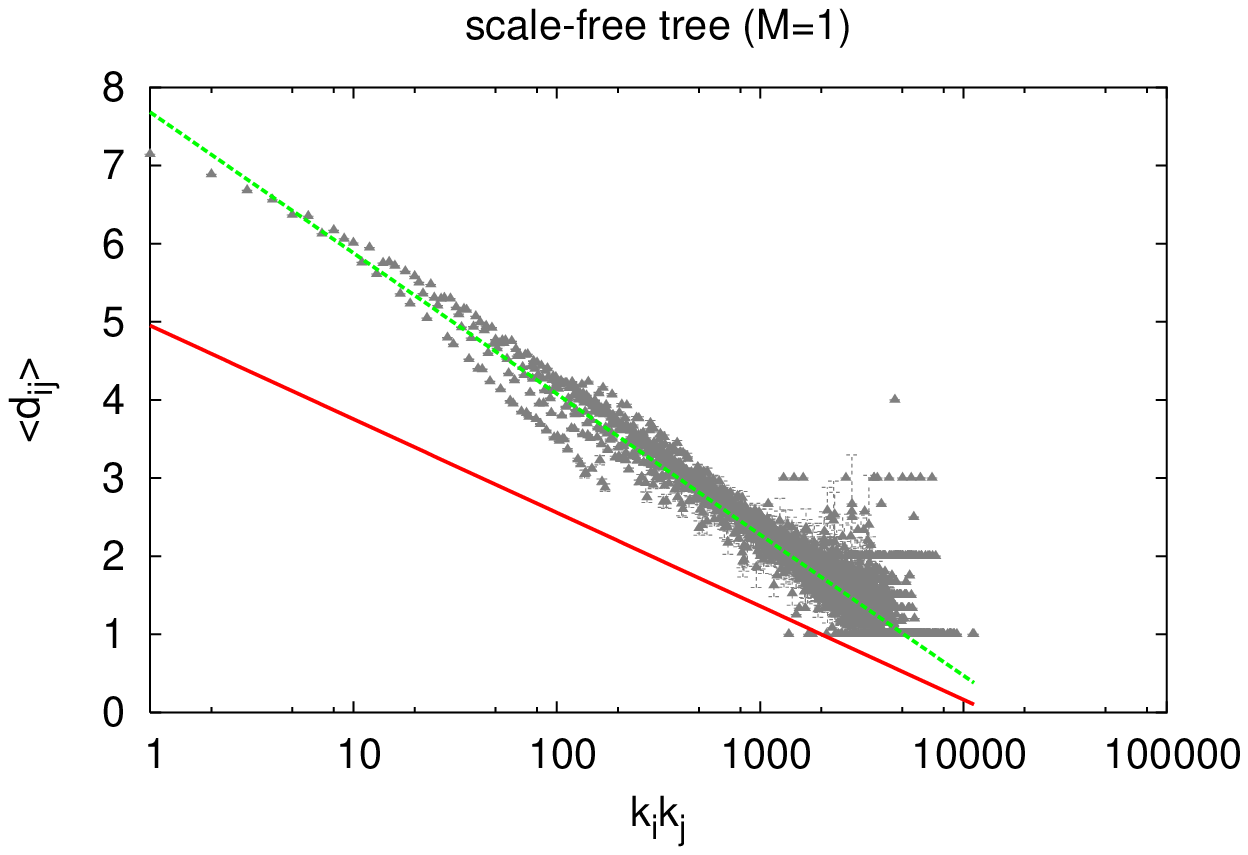}
\includegraphics[scale=.60]{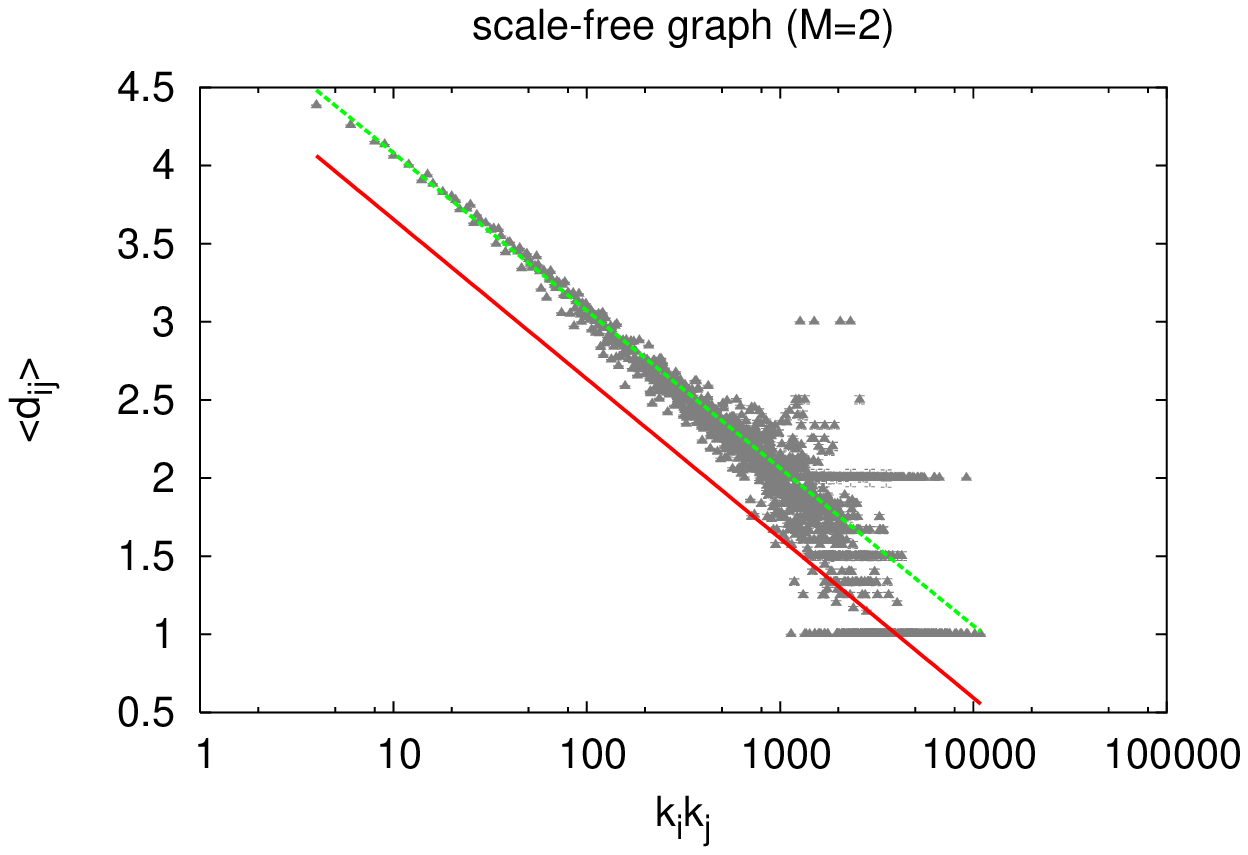}
\includegraphics[scale=.60]{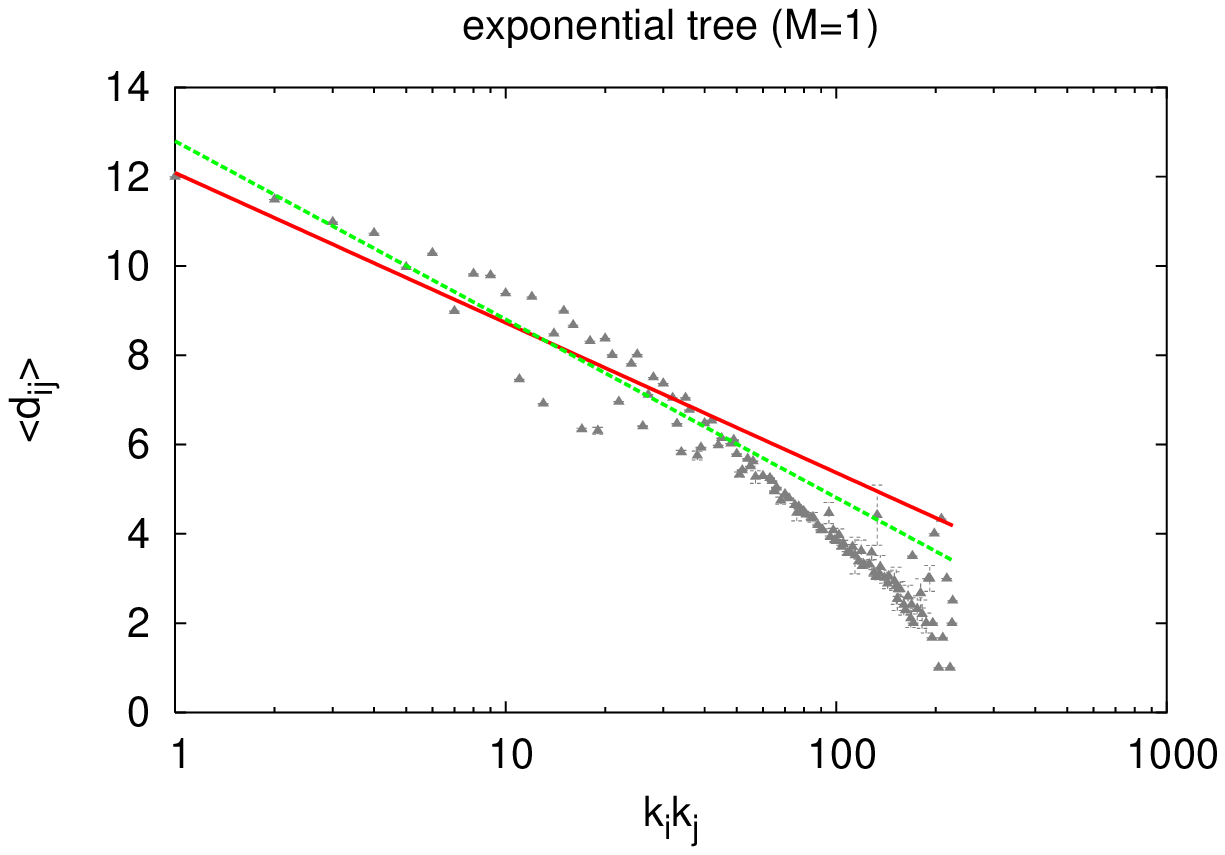}
\includegraphics[scale=.60]{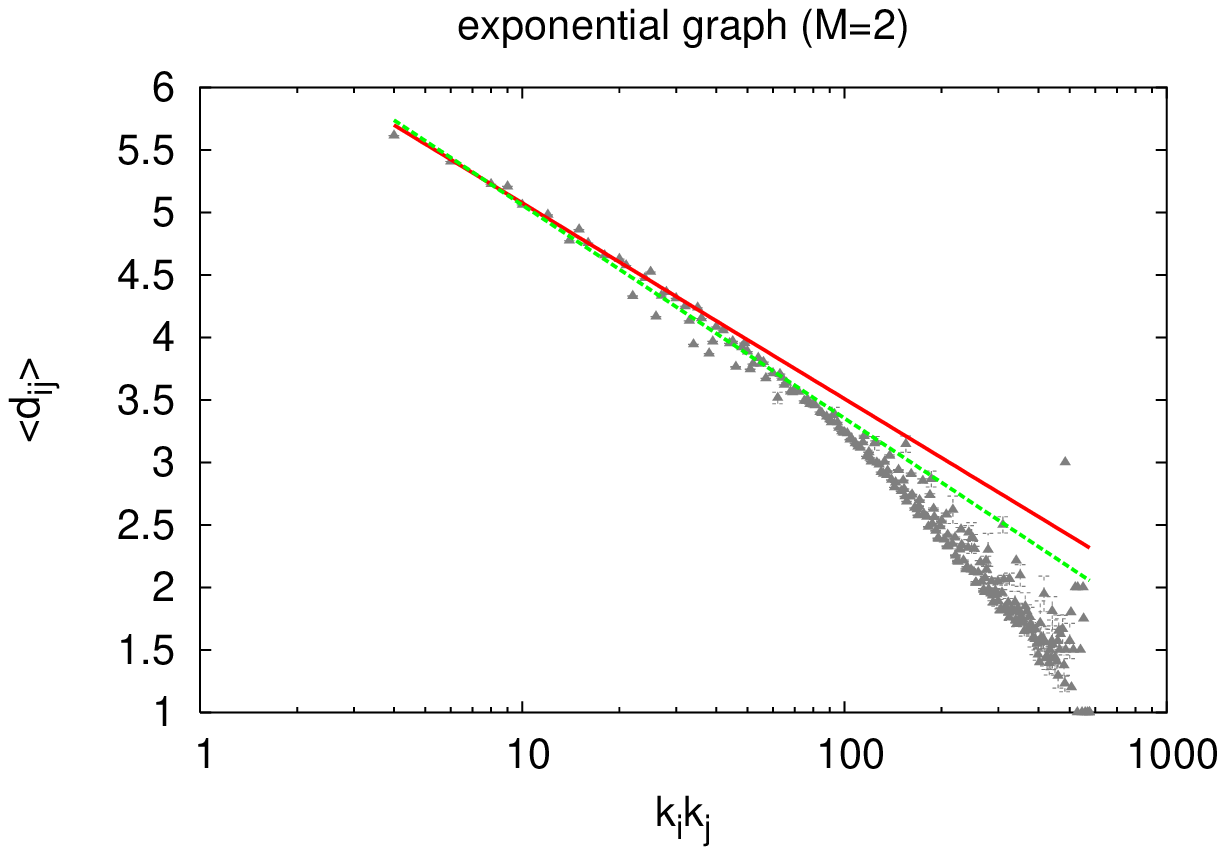}
\end{center}
\caption{\label{fig-dijvskikj} Dependence $\langle d_{ij}\rangle$ vs $k_i k_j$ for growing exponential and scale-free trees ($M=1$) and simple graphs ($M=2$).
The solid lines are the plots of Eq. \eqref{eq-motter} while the dotted lines result from a least-square fit.  $N=10^3$, $N_{\text{run}}=10^4$.}
\end{figure*}
%% ------------------------------------------------------

Here we show that this logarithmic dependence \eqref{eq-holyst} holds for exponential networks with $M=1,2$.  
The results are presented in Fig. \ref{fig-dijvskikj} and in Tab. \ref{tab}.
The least-square fit was confined to the two first decades of $k_ik_j$ values for the scale-free networks and to the one-and-half decade  of $k_ik_j$ values for the exponential ones.

%% ######################################################
\section{Discussion and conclusions}
%% ######################################################

Generating function mechanism \cite{newman,motter,shargel,ambjorn86,ambjorn90} has a mean-field nature and should work only for homogeneous trees.
This mechanism is founded upon the assumption that there are {\em no} correlations between nodes degrees.
But this assumption does not hold for growing (causal) trees, where the oldest nodes --- probably well connected --- are geometrically close \cite{burda-priva}.

However, the Motter {\em et al.} formula \eqref{eq-motter} works surprisingly nice also for growing networks, where triangles and other cyclic paths are possible.
The Motter {\em et al.} predictions agree with simulation particularly well for exponential networks, where also $z_m$ were reproduced quite fairly.
For scale-free networks this agreement is only qualitative.
It seems, that theoretical predictions given by Eqs. \eqref{eq-zm} and \eqref{eq-motter} (and obviously given by Eq. \eqref{eq-z2}) agree with results of the simulations for the networks for which degree distribution gives finite $\{k^2\}$.

Average number of vertexes in all generation $z_m$ is well known for homogeneous \cite{ambjorn86} and causal \cite{krapivsky01} trees.
The number $z_m$ of $m$-th neighbors derived in \cite{newman,motter} agree very well for small $m=3,4,5$ with the results of computer simulation for exponential networks where the old nodes, to which the new nodes are being attached, are chosen randomly.

Again, this should be valid for trees, but it works also nicely for $M=2$ when random attachment is used.
On the other hand the sum $\sum_{k=M}^\infty k(k-1)P_k(k)$ diverges for power-like distributions $P_k(k)$.
For finite but large lattices this sum ($\sigma$, Eq. \eqref{eq-sigma}) increases logarithmically with the system size $N$.
In all four investigated cases simulated number $z_2$ (given by number of ``2'' in distance matrix) agrees with $\{k^2\}-\{k\}$ (averaged over all graph nodes).

For larger $m$ formula \eqref{eq-zm} fails when applied to real networks, i.e. with finite $N$.
Usually, the second layer contains more nodes than the first one, from which follows that $z_2>z_1$.
Then --- accordingly to Eq. \eqref{eq-zm} --- $z_m$ increases with $m\in{\mathbb Z}$, but for finite systems it must start to decrease for large $m$.
In particular, any of $N$ nodes 
which constitute the network has no neighbors in $N$-th layer and does not posses any $N$-th 
neighbors ($z_m=0$ for $m\ge N$).
The distribution of the node-to-node distances for the growing networks discussed here were shown in 
\cite{task,app} and evaluated analytically in case of trees in Ref. \cite{burda}.

Still, the method of evaluation of $z_m$ ($m=3,4,5$) based on Eq. \eqref{eq-zm} may be quite useful.
The main effort should be paid to a theoretical evaluation of the average number of nodes in the second layer, i.e. the number of occurrences of ``2'' in the distance matrix, basing only on the degree distribution $P_k(k)$.
Such an evaluation of $z_2$ would allow, in principle, to reproduce the whole function $z_m$.

Although the node-to-node distance $\langle d_{ij}\rangle$ depends logarithmically on the product of the node $i$ and $j$ degrees (Fig. \ref{fig-dijvskikj}, Eq. \eqref{eq-holyst}, Ref. \cite{holyst}),
the dependence of the to-node distance on the node degree is not a trivial one \cite{epjb}.
We have demonstrated, that Eq. \eqref{eq-holyst} can be extended to the case of the growing exponential networks.
The Motter {\em et al.} predictions of values $A$ and $B$ in Eq. \eqref{eq-holyst} given by Eq. \eqref{eq-motter} agree for these networks quite fairly.

\newpage

%% ######################################################
\begin{acknowledgments}
%% ######################################################
Author thanks Krzysztof Ku{\l}akowski, Zdzis{\l}aw Burda and Andrzej Lenda for their valuable help and many fruitful discussions.
Calculations were carried out in ACK-CYFRONET-AGH.
The machine time on SGI~2800 is financed by the Polish Ministry of Science and Information Technology under grant No. KBN/\-SGI2800/\-AGH/\-018/\-2003.
\end{acknowledgments}

%% ######################################################

\end{document}